\documentclass[aps,prd,reprint,groupedaddress,onecolumn,showpacs]{revtex4-1}
\usepackage{graphicx}
\usepackage{bm}
\usepackage{amssymb,amsmath}

\newcommand{\be}{\begin{equation}}
\newcommand{\ee}{\end{equation}}
\newcommand{\ba}{\begin{array}}
\newcommand{\ea}{\end{array}}
\newcommand{\bqa}{\begin{eqnarray}}
\newcommand{\eqa}{\end{eqnarray}}

\begin{document}


\title{Hadron loops effect on mass shifts of the charmed and charmed-strange spectra}


\author{Zhi-Yong Zhou}
\email[]{zhouzhy@seu.edu.cn}
\affiliation{Department of Physics, Southeast University, Nanjing 211189,
P.~R.~China}
\author{Zhiguang Xiao}
\email[]{xiaozg@ustc.edu.cn}
\affiliation{Interdisciplinary Center for Theoretical Study, University of Science
and Technology of China, Hefei, Anhui 230026, China}


\date{\today}

\begin{abstract}
The hadron loop effect is conjectured to be important in understanding
discrepancies between the observed states in experiments and the
theoretical expectations of non-relativistic potential model. We
present that, in an easily operable procedure, the hadron loop effect
could shift the poles downwards to reduce the
differences and provide better descriptions of both the masses and the total widths, at least, of the radial quantum number $n=1$ charmed and charmed-strange
states. The $1^1P_1-1^3P_1$ mixing phenomena could be naturally
explained due to their couplings with common channels. The newly observed $D$ states are also addressed,
but there are still some problems remaining unclear.
\end{abstract}

\pacs{12.39.Jh, 13.20.Fc, 13.75.Lb, 11.55.Fv}

\maketitle
\section{Introduction}
Discoveries of more and more charmed or charmed-strange states in
experiments attract great interest on the theoretical side,
 because some members of them have unexpected properties.
 In the
Particle Data Group~(PDG)
table\cite{Nakamura:2010zzi}), six lower charmed states, $D^0$, $D^*(2007)^0$,
$D_0^*(2400)^0$, $D_1(2420)^0$, $D_1(2430)^0$, $D_2^*(2460)^0$, and
their partners have already been established. Recently, some evidences
of three new charmed states, $D(2550)$, $D(2610)$, and $D(2760)$ have
been reported by the BABAR Collaboration~\cite{delAmoSanchez:2010vq},
whose features lead to intense discussions and theoretical
suggestions of the further experimental
investigations~\cite{Sun:2010pg,Zhong:2010vq,Wang:2010ydc,Li:2010vx,Chen:2011rr,Guo:2011dd}.
There are also nine charmed-strange states quoted in the PDG table
among which some states' quantum numbers are undetermined.
The mass spectra of these charmed and charmed-strange states are roughly
depicted in the predictions of the non-relativistic potential model in the classic
work by Godfrey and Isgur~(referred as GI in the
following)~\cite{Godfrey:1985xj}. However, the observed masses are
generally lower than the predicted ones. For example, the biggest
discrepancies happening in both spectra are the $1^3P_0$ states. The $D_0^*(2318)$ is about $80\mathrm{MeV}$ lower than the expectation, while
the $D_{s0}^*(2317)$ is about $160\mathrm{MeV}$ lower. There are a body of theoretical
efforts at solving this
problem usually by changing the representation of the potential~(see Ref.~\cite{Isgur:1989vq,Ebert:2009ua,DiPierro:2001uu,Lakhina:2006fy,Matsuki:2007zza} and their references).
Lattice calculation has also been made to explain the
experiments\cite{Lewis:2000sv,Dougall:2003hv}. However, the present systematic
uncertainty of the Lattice calculations does not allow determinations of the charmed
mesons with a precision less than several hundred MeV.

Another expectation to shed light on this problem is to take the
coupled channel effects~(or called hadron loop effects) into account,
which plays an important role in understanding the enigmatic light
scalar spectrum and their decays~\cite{Zhou:2010ra,Tornqvist:1995kr}. In the light
scalar spectrum, the strong attraction of opened or closed channels
may dramatically shift the poles of the bare states to different
Riemann sheets attached to the physical region and the poles on
unphysical Riemann sheets appear as  peaks or just humps of the
modulus of scattering amplitudes in the experimental data. The mass
shifts induced by the intermediate hadron loops have also been
testified to present a better description of the charmonium
states~\cite{Heikkila:1983wd,Pennington:2007xr,Barnes:2007xu,Swanson:2006st}.  The
coupled channel effects have already brought some insights into the
nature of the charmed-strange $D_{sJ}(2317)$ and some other
states~\cite{vanBeveren:2003kd,Simonov:2004ar,Hwang:2004cd,Becirevic:2004uv}.
However, although this effect could explain some of the observed
charmed or charmed-strange states, there is still
a concern that
this effect may also exist in those states previously consistent with the
theoretical expectation~\cite{Godfrey:2005ww}. In this paper, we will
address this point by considering the mass shifts, induced by hadron loops, of all the firmly established charmed and charmed-strange
states. Here we propose an easily operable way, in which we use the
imaginary part of the self-energy funtion calculated from the quark
pair creation (QPC)
model~\cite{Micu:1968mk,Colglazier,LeYaouanc:1972ae} in the dispersion
relation to obtain the analytically continued inverse propagator and
extract the physical mass and width parameters, and then apply it to
the charmed and charmed-strange spectra to interpret their masses and
total decay widths in a consistent way. It is found that the results of their masses
and total widths are consistent with the experimental values, at least
for the non-radially-excited states. This picture gives a natural
explanation to the $1^1P_1-1^3P_1$ mixing by coupling with the same
channels instead of using a phenomenological mixing angle. This scheme has some similarities to the methods used by Heikkila $et\,
al.$~\cite{Heikkila:1983wd} and Pennington $et\ al.$ in their study of
the charmonium and bottomonium states, but there are significant
differences with them, as discussed in the text.

The paper is organized as follows: In Section II, the main scheme and
how to model the decay channels are briefly introduced. The mixing
mechanism is introduced in Section III. Numerical procedures and
results are discussed in Section IV. Section V is devoted to our
conclusions and further discussions.

\section{The scheme}

We start by considering a simple model at the hadron level, in which
the inverse meson propagator, $\mathbb{P}^{-1}(s)$, could be represented
as~\cite{Heikkila:1983wd,Pennington:2007xr}
\bqa\label{propagator}
\mathbb{P}^{-1}(s)=m_0^2-s+\Pi(s)=m_0^2-s+\sum_n\Pi_n(s),
\eqa
where $m_0$ is the mass of the bare $q\bar{q}$ state and $\Pi_n(s)$ is
the self-energy functions for the $n$-th decay channels. Here, the sum
is over all the opened channels or including nearby unopened
channels~(``just virtual").  $\Pi_n(s)$ is an analytic function with
only a right-hand cut starting from the $n$-th threshold $s_{th,n}$, and so one can write
its real part and imaginary part through a dispersion relation
\bqa\label{dr}
\mathrm{Re}\Pi_n(s)=\frac{1}{\pi}\mathcal{P}\int_{s_{th,n}}^\infty\mathrm{d}z\frac{\mathrm{Im}\Pi_n(z)}{(z-s)},
\eqa
where $\mathcal{P}\int$ means the principal value integration.
The pole of $\mathbb{P}(s)$ on the unphysical Riemann sheet attached
to the physical region specifies its mass and total width of a meson
by its position on the complex $s$ plane, usually defined as
$s_{pole}=(M_p-i\Gamma_p/2)^2$.

One could recover a generalization of the familiar Breit-Wigner
representation, usually used in experimental analyses, from
Eq.(\ref{propagator}), as
\bqa\label{BW}
\mathbb{P}^{-1}(s)=m(s)^2-s+im_{BW}\Gamma_{tot}(s),
\eqa
where $m(s)^2=m_0^2+\mathrm{Re}\Pi(s)$ is the ``runing squared mass"
and $\Gamma_{tot}(s)=\mathrm{Im}\Pi(s)/m_{BW}$. $m_{BW}$ is determined
at the real axis where $m(s)^2-s=0$ is fulfilled. The mass and width
parameters in these two definitions give similar results when one
encounters a narrow resonance, but they differ when the resonance is
board or when there are several poles interacting with each other.

\begin{figure}[h]%
\begin{center}%
\hspace{-2cm}\includegraphics[height=30mm]{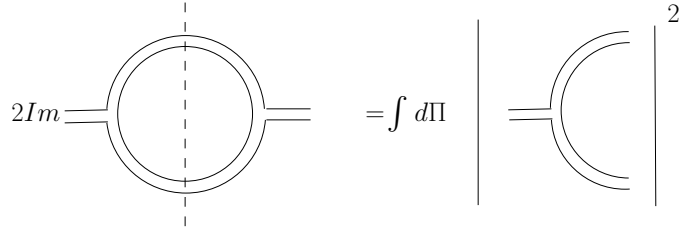}
\caption{\label{unitary} The imaginary part of the self-energy function. $\int\mathrm{d}\Pi$ means the integration over the phase space.}
\end{center}%
\end{figure}

Based on the Cutkosky rule, the imaginary part of the self-energy
function is expressed through the couplings between the bare state and
the coupled channels. The relation could be pictorially expressed as
Fig.\ref{unitary}.

Thus, one key ingredient of this scheme is to model the coupling
vertices in the calculation of the imaginary part of the self-energy
function. The QPC
model, also known as
the $^3P_0$ model in the literature, turns out to be applicable in
explaining the   Okubo-Zweig-Iizuka (OZI) allowed strong decays of a
hadron into two other hadrons, which are expected to be the dominant
decay modes of a meson if they are allowed. It is not only because
this model has proved to be successful but also because it could provide
analytical expressions for the vertex functions, which are convenient
for extracting the shifted poles in our scheme. Furthermore, the vertex functions have exponential factors which give a natural cutoff to the dispersion relation and we need not to choose one by hand as in Ref.~\cite{Pennington:2007xr}.

Here, we just make a brief review of the main results of the QPC
model used in our calculation.~(For a more complete review, see
\cite{yaouancbook,Blundell:1995ev,Luo:2009wu}) In the QPC model, the
meson~(with a quark $q_1$ and an anti-quark $q_2$) decay occurs by
producing a quark~($q_3$) and anti-quark~($q_4$) pair from the vaccum.
In the non-relativistic limit, the transition operator can be
represented as
\bqa
T=-3\gamma\sum_m\langle 1 m 1 -m|00\rangle\int d^3\vec{p_3}d^3\vec{p_4}\delta^3(\vec{p_3}+\vec{p_4})\mathcal{Y}_1^m(\frac{\vec{p_3}-\vec{p_4}}{2})\chi_{1 -m}^{34}\phi_0^{34}\omega_0^{34}b_3^\dagger(\vec{p_3})d_4^\dagger(\vec{p_4}),
\eqa
where $\gamma$ is a dimensionless model parameter and
$\mathcal{Y}_1^m(\vec{p})\equiv p^lY_l^m(\theta_p,\phi_p)$ is a solid
harmonic that gives the momentum-space distribution of the created
pair. Here the spins and relative orbital angular momentum of the
created quark and anti-quark~(referred to by subscripts $3$ and $4$,
respectively) are combined to give the pair the overall
$J^{PC}=0^{++}$ quantum numbers.
$\phi_0^{34}=(u\bar{u}+d\bar{d}+s\bar{s})/\sqrt{3}$ and
$\omega_0^{34}=\delta_{ij}$, where $i$ and $j$ are the SU(3)-color
indices of the created quark and anti-quark. $\chi_{1 -m}^{34}$ is a
triplet of spin.

Define the $S$ matrix for the meson decay $A\rightarrow BC$ as
\bqa
\langle BC|S|A\rangle=I-2\pi i \delta(E_f-E_i)\langle BC|T|A\rangle,
\eqa
and then
\bqa
\langle BC|T|A\rangle=\delta^3(\vec{P_f}-\vec{P_i})\mathcal{M}^{M_{J_A}M_{J_B}M_{J_C}}.
\eqa
The amplitude turns out to be
\bqa
\mathcal{M}^{M_{J_A}M_{J_B}M_{J_C}}(\vec{P})&=&\gamma\sqrt{8E_AE_BE_C}\sum_{M_{L_A},M_{S_A},M_{L_B},M_{S_B},M_{L_C},M_{S_C},m}\langle L_A M_{L_A}S_A M_{S_A}|J_A M_{J_A}\rangle\nonumber\\
&&\times\langle L_B M_{L_B}S_B M_{S_B}|J_B M_{J_B}\rangle\langle L_C M_{L_C}S_C M_{S_C}|J_C M_{J_C}\rangle \langle 1 m 1 -m|00\rangle\nonumber\\
&&\times \langle \chi_{S_C M_{S_C}}^{32}\chi_{S_B M_{S_B}}^{14}|\chi_{S_A M_{S_A}}^{12}\chi_{1 -m}^{34}\rangle \langle \phi_C^{32}\phi_B^{14}|\phi_A^{12}\phi_0^{34}\rangle I_{M_{L_B},M_{L_C}}^{M_{L_A},m}(\vec{P}).
\eqa
The spatial integral $I_{M_{L_B},M_{L_C}}^{M_{L_A},m}(\vec{P})$ is given by
\bqa
I_{M_{L_B},M_{L_C}}^{M_{L_A},m}(\vec{P})=\int d^3\vec{k}\psi^*_{n_B
L_B M_{L_B}}(-\vec{k}+\frac{\mu_4}{\mu_1+\mu_4}\vec{P})\psi^*_{n_C L_C
M_{L_C}}(\vec{k}-\frac{\mu_3}{\mu_2+\mu_3}\vec{P})\psi_{n_A L_A
M_{L_A}}(-\vec{k}+\vec{P})\mathcal{Y}_1^m(\vec{k}),
\eqa
where we have taken $\vec{P}\equiv \vec{P_B}=-\vec{P_C}$ and $\mu_i$
is the mass of the $i$-th quark. $\psi_{n_A L_A M_{L_A}}(\vec{k_A})$
is the relative wave function of the quarks in meson $A$ in the
momentum space.

The recoupling of the spin  matrix element can be written, in terms of
the Wigner's $9$-$j$ symbol, as~\cite{yaouancbook}
\bqa
\langle \chi_{S_C M_{S_C}}^{32}\chi_{S_B M_{S_B}}^{14}|\chi_{S_A M_{S_A}}^{12}\chi_{1 -m}^{34}\rangle =[3(2S_B+1)(2S_C+1)(2S_A+1)]^{1/2}\nonumber\\ \times\sum_{S,M_S}\langle S_C M_{S_C}S_B M_{S_B}|S M_S\rangle \langle S M_S|S_A M_{S_A};1,-m\rangle
\left\{
\begin{array}{ccc}
                                                                                                1/2 & 1/2 & S_C \\
                                                                                                1/2 & 1/2 & S_B \\
                                                                                                S_A & 1 & S \\
                                                                                          \end{array}
											  \right\}.
\eqa
The flavor matrix element is
\bqa
\langle\phi_C^{32}\phi_B^{14}|\phi_A^{12}\phi_0^{34}\rangle=\sum_{I,I^3}\langle I_C,I_C^3;I_B I_B^3|I_A I_A^3\rangle[(2I_B+1)(2I_C+1)(2I_A+1)]^{1/2}
\left\{\begin{array}{ccc}
                                                                                                I_2 & I_3 & I_C \\
                                                                                                I_1 & I_4 & I_B \\
                                                                                                I_A & 0 & I_A \\
                                                                                              \end{array}
                                                                                            \right\},
\eqa
where $I_i (I_1, I_2, I_3, I_4)$ is the isospin of the quark $q_i$.

The imaginary part of the self-energy function in the dispersion
relation, Eq.(\ref{dr}), could be expressed as
\bqa\label{discont}
Im\Pi_{A\rightarrow BC}(s)=-\frac{\pi^2}{2J_A+1} \frac{|\vec{P}(s)|}{\sqrt{s}}\sum_{M_{J_A},M_{J_B},M_{J_C}}|\mathcal{M}^{M_{J_A},M_{J_B},M_{J_C}}(s)|^2,
\eqa
where $|P(s)|$ is the three momentum of $B$ and $C$ in their center of mass frame.  So,
\bqa
\frac{|P(s)|}{\sqrt{s}}=\frac{\sqrt{(s-(m_B+m_C)^2)(s-(m_B-m_C)^2)}}{2s}.
\eqa
Care must be taken when Eq.(\ref{discont}) is continued to the complex
$s$ plane. Since what is used in this model is only the tree level
amplitude, there is no right hand cut for
$\mathcal{M}^{M_{J_A},M_{J_B},M_{J_C}}(s)$. Thus, the analytical
continuation of the amplitude obeys
$\mathcal{M}(s+i\epsilon)^*=\mathcal{M}(s-i\epsilon)=\mathcal{M}(s+i\epsilon)$.
The physical amplitude  with loop contributions should have
right hand cuts, and, in principle, the analytical continuation turns
to be
$\mathcal{M}(s+i\epsilon)^*=\mathcal{M}(s-i\epsilon)=\mathcal{M}^{n}(s+i\epsilon)$
by meeting the need of real analyticity.
$\mathcal{M}(s+i\epsilon)$ means the amplitude on the physical Riemann sheet~(the first sheet, in language of the analytic $S$ matrix theory), and $\mathcal{M}^{n}(s+i\epsilon)$ means the amplitude on the unphysical
Riemann sheet~(the $n$-th sheet) attached with the physical region.

With the analytical expression of the imaginary part of the coupled
channel, one will be able to extract the poles or the Breit-Wigner
parameters from the propagators by standard procedures. In principle,
all hadronic channels should contribute to the meson mass, as considered by  Heikkila $et\
al.$ in studying the charmonium states~\cite{Heikkila:1983wd}. Even all the ``virtual" channels will
contribute to the real parts of $\Pi(s)$ and renormalize the ``bare"
mass. Pennington $et\
al.$ proposed that a once-subtracted dispersion relation will suppress contributions of the faraway ``virtual" channels and make the picture simpler~\cite{Pennington:2007xr}.  Since what we consider here is only the mass shifts, we could
make a once-subtracted dispersion relation at some suitable point
$s=s_0$. It is reasonably expected that the lowest
charmed state, $D^0$, as a bound state, has the mass defined by the
potential model, uninfluenced by the effect of the hadron loops. Its
mass then essentially defines the mass scale and thus fixes the
subtracted point. So, we set the subtracted point $s_0=m_{D^0}^2$ or
$s_0=(m_c+m_u)^2$ in a practical manner. The inverse of the meson
propagator turns out to be
\bqa
\mathbb{P}^{-1}(s)=m_{pot}^2-s+\sum_n\frac{s-s_0}{\pi}\int_{s_{th,n}}^\infty\mathrm{d}z\frac{\mathrm{Im}\Pi_n(z)}{(z-s_0)(z-s)},
\eqa
where $m_{pot}$ is the bare mass of a certain meson defined in the potential model.

\section{Mixing mechanism}

In this scheme, all the states with the same spin-parity have
interference effects and could mix with each other. For example, the
two $J = 1$ states of the $P$-wave are usually regarded as
linear combinations of $^3P_1$ and $^1P_1$ assignments.
Here in considering the coupled channel effect, the mixing mechanism
comes from the coupling via common channels. It is also believed that
the $2^3S_1$ and $1^3D_1$ states mix with each other, similar to the
interpretation of the charmonium $\psi(3770)$
state~\cite{Rosner:2004wy}.

The inverse of the propagator with two bare states mixing with each
other reads
\bqa
\mathbb{P}^{-1}(s)=\left(
                     \begin{array}{cc}
                       M_{11}^2(s) & M_{12}^2(s) \\
                       M_{12}^2(s) & M_{22}^2(s) \\
                     \end{array}
                   \right)-\delta_{a,b}s
=\left(
   \begin{array}{cc}
     m_{bare,1}^2-s+\Pi_{11}(s) & \Pi_{12}(s) \\
     \Pi_{21}(s) & m_{bare,2}^2-s+\Pi_{22}(s) \\
   \end{array}
 \right)
,
\eqa
where $M_{a,b}^2(s)$ is  the mass matrix and $m_{bare,a}$ represents
the mass parameter of the bare $a$ state. The off-diagonal terms of
the self-energy function is represented by the $1PI$ diagram for the
two mixed states. The physical states should be determined by the
meson propagator matrix after diagonalization
\bqa
 M_{diag}^2(s)=\alpha(s)^{-1} M_{a,b}^2(s)\alpha(s),
\eqa
where the mixing matrix $\alpha(s)$ satisfies $\alpha(s)^T
\alpha(s)=I$, $i.e.$, $\alpha(s)$ is a complex orthogonal matrix since
$M^2_{a,b}(s)$ is symmetric. The $\alpha(s)$ matrix turns to be
complex when the thresholds are open.  The physical poles could be
extracted, in an equivalent way, by finding the zero points of the
determinant of the inverse propagator, that is to solve
$det(\mathbb{P}^{-1}(s))=0$.

\section{Numerical analyses}

The bare masses of the related mesons are chosen at the values
of the GI's work~\cite{Godfrey:1985xj}, as also listed in
Table.\ref{compilation} for comparison. As for the dimensionless
parameter, $\gamma$, and the effective $\beta$ parameters in the QPC
model to characterize the harmonic oscillator wave functions, we
choose the same values as determined from the potential in GI's work
for self-consistency. The constituent quark masses are $M_c=1.628\mathrm{GeV}$,
$M_s=0.419\mathrm{GeV}$, and $M_u=0.22\mathrm{GeV}$. $\gamma=6.9$ and the
values of $\beta$s are from Ref.\cite{Godfrey:1986wj,Kokoski:1985is}.
The physical masses concerned in the final states are the average values
 in the PDG table. The relative wave functions between the
quarks in the mock-meson states are simple harmonic oscillator~(SHO)
wave functions usually used in the QPC model calculation, which brings
some uncertainties into the calculation, as discussed later.

There are some further explanations for the effective $\beta$
parameters of $c\bar{u}$ states. Godfrey and Isgur have only presented
their results of $n=1$ $S$ and $P$-wave charmed states but not
provided
those of the $D$-wave and radial excited states which is
needed in our discussion of the newly observed charmed states. We can only estimate the values by
assuming that their ratios between the $\beta$ values of the $c\bar{u}$
states are similar to the ratios in the results from the other research
groups. For example, we find the ratios of the $\beta$ values between
different charmed states in Ref.\cite{Close:2005se} and
Ref.\cite{Li:2010vx} are almost same. Thus, the $\beta$ values of
$c\bar{u}$ states used in our calculation, except those listed in
Ref.\cite{Kokoski:1985is}, are $\beta(1^3D_j)=0.44\pm
0.02\mathrm{GeV}$, $\beta(2^1S_0)=0.47\pm 0.02\mathrm{GeV}$, and
$\beta(2^3S_1)=0.44\pm 0.02\mathrm{GeV}$, respectively.

The opened or nearby unopened two-body channels taken into account in our calculations are all
listed in Table.\ref{decaychannel}. Those channels with the $\sigma$ meson are
not considered,
because $\sigma$ meson is not regarded as a conventional $q\bar{q}$ state in
the potential model~\cite{Godfrey:1985xj}. It will quickly decay into two pions
and the three-body decays usually present a minor contribution.

\begin{table}[htdp]
\begin{center}
\begin{tabular}{c c c c c c c c c c c c}
\hline
Mode & channel &$1^-(1^3S_1)$& $0^+(1^3P_0)$ & $1^+(1^1P_1)$ & $1^+(1^3P_1)$ & $2^+(1^3P_2)$ & $0^-(2^1S_0)$&$1^-(2^3S_1)$&$1^-(1^3D_1)$&$3^-(1^3D_3)$ \\
\hline
$0^-+0^-$ & $D\pi$ &$\Box$ & $\Box$ &  &   &$\Box$ & &$\Box$ &$\Box$ &$\Box$ \\
         & $D\eta$ & &  &  &  & & & $\Box$&$\Box$ &$\Box$ \\
         & $D_sK$ & &  &  &  & & &$\Box$ & $\Box$&$\Box$ \\
$1^-+0^-$ & $D^*\pi$ & &  & $\Box$ & $\Box$ & $\Box$&$\Box$ &$\Box$ &$\Box$&$\Box$ \\
  & $D^*\eta$ &  &  &  &  & & & $\Box$&$\Box$ &$\Box$ \\
  & $D_s^*K$ & &  &  &  & & & &$\Box$ &$\Box$ \\
$0^-+1^-$ & $D\rho$ & &  &  &  & & & &$\Box$ &$\Box$ \\
          & $D\omega$ & &  &  &  & & & &$\Box$ &$\Box$ \\
$0^++0^-$ & $D_0^*\pi$ & &  &  &  & & $\Box$&& & \\
$1^+(T)+0^-$ & $D_1(2420)\pi$ & &  &  &  & & & $\Box$&$\Box$ &$\Box$ \\
$1^+(S)+0^-$ & $D_1(2430)\pi$ & &  &  &  & & &$\Box$ & $\Box$&$\Box$ \\
$2^++0^-$ & $D_2^*(2460)\pi$ & &  &  &  & & &$\Box$ & $\Box$&$\Box$ \\
\hline
\end{tabular}
\caption{\label{decaychannel}The channels of the charmed states considered in this paper.}
\end{center}
\end{table}%

The masses and widths are simultaneously determined,
as listed in Table.\ref{compilation}, for the charmed states, where we present the pole positions
as well as the Breit-Wigner parameters for comparison. Remarkable improvements of the shifted masses of the
already established charmed mesons could be found instantly. Furthermore, the
total widths specified by twice of the imaginary part of the pole positions are
also consistent in good quality with the values in the PDG table.
\begin{table}[htdp]
\begin{center}
\begin{tabular}{|c|c|c|c|c|c|c|}
\hline
$J^{P}(n^{2s+1}L_J)$& Expt. mass&Expt. width & $m_{BW}$&$\Gamma_{BW}$& $\sqrt{s_{pole}}=M-i\Gamma/2$ & GI mass  \\
\hline
$0^-(1^1S_0)$ & 1867 &       &         &&     & 1880  \\
$1^-(1^3S_1)$ & $2007\pm 0.16$ & $<2.1$ &2016&0.02& $2016-0.01 i$&2040\\
$0^+(1^3P_0)$ & $2318\pm 29$ &$267\pm 40$&2335&233    &$2275-125 i$  & 2400  \\
$``1^+(1^1P_1)"$ & $2422\pm 0.6$ &$20\pm 1.7$&2420&16     &$2410-7 i$   & 2440 \\
$``1^+(1^3P_1)"$ & $2427\pm 40$ &$384^{+130}_{-110}$&2409&163     &$2377-94 i$ & 2490 \\
$2^+(1^3P_2)$ & $2462\pm 1$ &$43\pm 3$&2453& 26   & $2452-12 i$  & 2500 \\
\hline
$0^-(2^1S_0)$ & $2533\pm 14(?)$ &$128\pm 33(?)$ &2534&25     &$2533-12 i$ & 2580 \\
$1^-(2^3S_1)$ & $2608\pm 5(?)$ & $93\pm 19(?)$&2525&8 &$2523-5 i$    & 2640 \\
$1^-(1^3D_1)$ &$ 2763\pm 5(?)$ & $61\pm 9(?)$ &2730&120&$2686-66 i$    & 2820 \\
$3^-(1^3D_3)$ & (?) &(?)&2735&9&$2735-4 i$          & 2830 \\
\hline
\end{tabular}
\end{center}
\caption{\label{compilation}Compilation of the experimental masses and the total widths(the PDG average values~\cite{Nakamura:2010zzi}) of the charmed states, the shifted pole positions and the mass spectrum in the GI's model~\cite{Godfrey:1985xj}. The experimental values of $2^1S_0$, $2^3S_1$, and  $1^3D_1$ are from Ref.\cite{delAmoSanchez:2010vq}. Here we only list the neutral charmed states. The unit is $\mathrm{MeV}$.}
\end{table}%

The $D(1^1S_0)$ state is a long-lived particle in the strong interaction and
there is no opened strong channel, so we regard it to be well described as a
bound state in the potential model and  choose its squared mass as the
subtraction point of the dispersion relations.

 When we calculate the mass shift
of the $D(1^3S_1)$,  the $D^0\pi^0$ and $D^+\pi^-$ threshold are both taken
into account, because the $D^+\pi^-$ threshold is at about $2009\mathrm{MeV}$,
just  $2\mathrm{MeV}$ higher than the observed $D^*(2007)^0$. It is a
typical ``just virtual" channel and in principle it will contribute a
significant mass shift to the bare state. If the coupling to the
$D^+\pi^-$ threshold is excluded, the pole mass will only be shifted
to about $2031MeV$ using this
set of parameters.

The pole of $D(1^3P_0)$ is significantly shifted down to $2275\mathrm{MeV}$,
which is $125\mathrm{MeV}$ down below the potential model prediction. The pole
width is about $250\mathrm{MeV}$, which is in accordance with the experimental
value within errorbar.

The shifted pole mass favors the BABAR and Belle
results~\cite{Aubert:2009wg,Abe:2003zm} over the FOCUS
result~\cite{Link:2003bd}. The $D(1^1P_1)$ and $D(1^3P_1)$ states stay close to
each other and they both have the same quantum numbers $J^{P}=1^+$ and similar
decay channels. The unmixed pole positions are at
$\sqrt{s_{(1^1P_1)}}=2387-i28\mathrm{MeV}$ and
$\sqrt{s_{(1^3P_1)}}=2427-i71\mathrm{MeV}$, respectively. Both of the masses
and the widths of the unmixed $1^1P_1$ state have large differences from the
experimental values. It is the effect of their couplings with common channel $D^*\pi$
that significantly change their pole positions to one narrower and the other
boarder. The poles determined by the zero points of the inverse propagator
matrix are at $\sqrt{s_{(``1^1P_1")}}=2410-i7 \mathrm{MeV}$ and
$\sqrt{s_{(``1^3P_1")}}=2377-i94 \mathrm{MeV}$ respectively, which characterize
the two observed states quite well. Their related Breit-Wigner parameters agree with the experimental values better. It is interesting to mention that in this
scheme the mixing matrix, as a function of $s$, are complex-valued and it is not easy
to find its relation with the mixing angle commonly used in phenomenological
analyses. Here, the approximate values of the mixing matrix at about
$2410\mathrm{MeV}$ is
\bqa
\alpha|_{s=(2.41\mathrm{GeV})^2}=\left(
  \begin{array}{cc}
   -0.57+0.26 i & 0.87+0.17 i \\
     0.87+0.17 i &0.57-0.26 i \\
  \end{array}
\right),
\eqa
whose imaginary parts are small compared with the real parts. If one just
neglect the imaginary parts and define the mixing matrix as usual, one obtains
\bqa
\left(
  \begin{array}{cc}
   -\mathrm{sin}\theta &  \mathrm{cos}\theta \\
    \mathrm{ cos}\theta & \mathrm{sin}\theta \\
  \end{array}
\right)\simeq\left(
  \begin{array}{cc}
   -0.57 & 0.87 \\
     0.87 &0.57 \\
  \end{array}
\right),
\eqa
which means the mixing angle $\theta\simeq(29^\circ\sim 35^\circ)$. It is in
agreement with the value $\theta\simeq 35.3^\circ$ obtained by considering the
heavy quark symmetry~\cite{Barnes:2002mu}.

As for the other charmed states which are not quoted in the PDG table, the
evidences of several new charmed states, $D(2550)$, $D(2610)$, and $D(2760)$,
have been recently reported by the BABAR
collaboration~\cite{delAmoSanchez:2010vq}. Several groups have presented their
tentative interpretations of the nature of these
states~\cite{Sun:2010pg,Zhong:2010vq,Wang:2010ydc,Li:2010vx,Chen:2011rr}, and
we make a brief summary of their conclusions here. $D(2550)$ is assigned to the
$2^1S_0$, but its large decay width could not be explained by the QPC model,
the chiral quark model, and the relativistic quark model, so further
experimental explorations were suggested. Although the potential model has
predicted $D(2^3S_1)$ to be located at about $2640\mathrm{MeV}$, the QPC model and
the chiral quark model also favor  $D(2610)$ to be a mixed state of
$D(2^3S_1)$ and $D(1^3D_1)$ to interpret its large width. There are conflicting
opinions about the assignment of  $D(2760)$ as the heavier mixed state of
$D(2^3S_1)$ and $D(1^3D_1)$, or as $D(1^3D_3)$.

In our calculation, the mass shift induced by the intermediate states also
reduces the pole masses of $D(2^1S_0)$, down to $2533\mathrm{MeV}$, but its pole
width is quite narrow compared with the experimental value, as shown in
Table.\ref{compilation}. However, the mass of $D(2^3S_1)$ is shifted too much
down to about $2523\mathrm{MeV}$ due to many intermediate channels opened.
Actually, the pole is even shifted down about $100\mathrm{MeV}$ below some
thresholds, and its pole width is fairly small as well. It seems to become a
quasi-bound state due to its strong coupling with the $D_1\pi$ channel.  Of
course, there is still some parameter space for the effective $\beta$
parameter and the dimensionless coupling strength parameter $\gamma$ to be
tuned to reduce the mass shift to fit the experiment signal, because these
parameters have significant uncertainties. In our opinion, one
possible reason why the result seems to be inaccurate is the uncertainty of the
SHO function we used to estimate the coupling vertices, which might have a
larger tail than the realistic one in the high $s$ region, which will contribute to
the mass shift through the dispersion relation.  A more realistic wave function
solved from the linear potential model could be more reliable to describe
the meson property. However, usually, this kind of wave function does not have
an analytical representation and it can not be easily continued into the complex
$s$ plane in our scheme. On the other hand, it is too early to get any firm
conclusion, since these states still need further experimental confirmations.
The pole of $D(1^3D_1)$ is shifted down to $2686-i66\mathrm{MeV}$ as well,
whose Breit-Wigner mass is about $2735\mathrm{MeV}$ which is closer to the mass of
$D(2760)$. Unlike the $1^1P_1-1^3P_1$ case, with this set of parameters, the
mixing mechanism due to the coupling with their common channels does not change their
positions much. The unmixed pole of $D(1^3D_1)$ could also be estimated at
about $2735$ but its width is narrow.
\begin{table}[htdp]
\begin{center}
\begin{tabular}{c c c c c c c c c c c c}
\hline
Mode & channel &$1^-(1^3S_1)$& $0^+(1^3P_0)$ & $1^+(1^1P_1)$ & $1^+(1^3P_1)$ & $2^+(1^3P_2)$  \\
\hline
$0^-+0^-$ & $DK$ &$\Box$ & $\Box$ &  &   &$\Box$  \\
$1^-+0^-$ & $D^*K$ & &  & $\Box$ & $\Box$ & $\Box$\\
\hline
\end{tabular}
\caption{\label{csdecaychannel}The opened and nearby closed channels of the charmed-strange states considered in this paper.}
\end{center}
\end{table}%

\begin{table}[htdp]
\begin{center}
\begin{tabular}{|c|c|c|c|c|c|c|}
\hline
$J^{P}(n^{2s+1}L_J)$& Expt. mass&Expt. width & $\sqrt{s_{pole}}=M-i\Gamma/2$ & GI mass  \\
\hline
$0^-(1^1S_0)$ & 1968 &       &            & 1980  \\
$1^-(1^3S_1)$ & $2112\pm 0.5$ & $<1.9$ & $2114-0 i$&2130\\
$0^+(1^3P_0)$ & $2317\pm 0.6$ &$<3.8$&$2358-0 i$  & 2480  \\
$``1^+(1^1P_1)"$ & $2459\pm 0.6$ &$<3.5$&$2470-0 i$   & 2530 \\
$``1^+(1^3P_1)"$ & $2535\pm 0.2$ &$<2.3$ &$2508-1 i$ & 2570 \\
$2^+(1^3P_2)$ & $2573\pm 1$ &$20\pm 5$  & $2522-7 i$  & 2590 \\
\hline
\end{tabular}
\end{center}
\caption{\label{cscompilation}Compilation of the experimental masses and the total widths(the PDG average values~\cite{Nakamura:2010zzi} of the charmed strange states. The unit is $\mathrm{MeV}$.}
\end{table}%

The discrepancies that happen in the charmed-strange spectrum could be
well addressed qualitatively, owing to their coupling with the opened
thresholds and the nearby unopened OZI-allowed strong thresholds in
Table.\ref{csdecaychannel}, as the picture proposed by van Beveren and
Rupp for explaining the $D_{sJ}^*(2317)$ state
\cite{vanBeveren:2003kd}. Some the thresholds are opened as a
result of the isospin breaking effects, $e.g.$,
$D_{sJ}^*(2317)\rightarrow D_s\pi^0,D_s^*\pi^0$, whose contributions
are highly suppressed by a factor of about
$(m_u-m_d)/(m_s-(m_u+m_d)/2)\approx 1/38$, where the masses are the
current quark masses. The coupling to such thresholds will
contribute tiny imaginary parts of the self-energy functions, which
hardly shift the mass of the state and only contribute to the decay
widths with an order of
$\mathrm{KeV}$. So we completely neglect these thresholds with
isospin breaking effects and those OZI suppressed. When we choose the
value of dimensionless strength parameter  $\gamma=6.9$ for the
charmed-strange spectrum, the mass shifts will be a little larger.
We change $\gamma$ to be around $5.5$ and obtained the shifted
masses of the charmed-strange $S$ and $P$-wave states, as listed in
Table.\ref{cscompilation}. One could regard this fine tuning procedure
as a ``fit", because we only want to give a qualitative description
for the charmed-strange spectrum. Indeed the $\gamma$ parameter in the QPC model, determined by fitting to experimental decay processes, usually has an uncertainty of about
$30\%$~\cite{Close:2005se,Li:2010vx}. Here we
only list the pole positions, because they do not differ much with the
Breit-Wigner parameters in this case as quasi-bound states or narrow
states.

If the isospin breaking and other weak interaction channels are
unopened, $D_{s}^*(2112)$ and $D_{sJ}^*(2317)$ are the bound states when
the bare $1^1S_0$ and $1^3S_1$ states are coupled to the ``just virtual" $DK$
threshold. They show as the poles on the real axis of the physical Riemann sheet.
It is the coupling of the bare states to the lower
isospin breaking $D_s\pi^0$ thresholds and the other weak thresholds
that
shift the poles to the unphysical Riemann sheets when they are open in
reality. When the mixing of the $1^1P_1$ and $1^3P_1$ states is not
considered, they are both the bound states at $2478\mathrm{MeV}$ and
$2493\mathrm{MeV}$ respectively below the $D_s^*K$ thresholds. The
mixing owing to coupling with the common $D_s^*K$ thresholds shifts
the $1^1P_1$ downwards along the real $s$ axis, and the $1^3P_1$ moves
upwards and crosses $D_s^*K$ thresholds into the complex $s$ plane of
unphysical Riemann sheet.

\section{Conclusions}

In this paper, we propose a simple procedure to extract the pole
positions or determine the Breit-Wigner parameters of the charmed
states based on the parameters in the non-relativistic potential
model, by using the analytical representation of the QPC model to
mimic the behaviors of the imaginary part of the self-energy function
of the meson propagator. Overall improvements could be found
between the pole positions or Breit-Wigner parameters and the
well-established charmed and charmed-strange mesons. Several
charmed-strange states could be regarded as the quasi-bound states,
due to the coupling with nearby unopened OZI-allowed thresholds. In
this model, the $1^1P_1-1^3P_1$ mixing is explained by the coupling
with common channels and these resultant pole masses and widths are
consistent with the observed values. It is worth stressing that our calculation is the first one that systematically addresses such a broad spectrum and the decays of the members by considering the coupled channel effects, as far as we know. This calculation may help to
improve our understanding the charmed and charmed-strange spectra.

There are still some differences between the shifted pole positions
and the parameters of the newly observed states. Since at the present
stage the statistics of the data is still not enough to make a firm
determination, further experimental evidences are required for a
confirmation of these mesons.

\begin{acknowledgments}
  We are grateful to valuable discussions about the details of the QPC model with Xiang Liu and Zhi-Feng Sun.  ZX thanks
Chinese Universities Scientific Fund for support. It is partly supported by China National Natural
Science Foundation under contract No.10705009 and No. 10875001.
\end{acknowledgments}

\bibliographystyle{apsrev4-1}
\bibliography{charmed}

\end{document}